\documentclass{article}

\def\ignore#1{}

\def\psfancypar#1#2{\begingroup\def\par{\endgraf\endgroup\lineskiplimit=0pt}
               \setbox2=\hbox{\large\sc #2}
               \newdimen\tmpht \tmpht \ht2 \advance\tmpht by \baselineskip
               \font\hhuge=Times-Bold at \tmpht
               \setbox1=\hbox{{\hhuge #1}}
               \count7=\tmpht \count8=\ht1
               \divide\count8 by 1000 \divide\count7 by \count8 
               \tmpht=.001\tmpht\multiply\tmpht by \count7 
               \font\hhuge=Times-Bold at \tmpht
               \setbox1=\hbox{{\hhuge #1}}
               \noindent
                \hangindent1.05\wd1
               \hangafter=-2 {\hskip-\hangindent
               \lower1\ht1\hbox{\raise1.0\ht2\copy1}%
                \kern-0\wd1}\copy2\lineskiplimit=-1000pt}

\def\boxit#1{\vbox{\hrule\hbox{\vrule\kern3pt
        \vbox{\kern3pt#1\kern3pt}\kern3pt\vrule}\hrule}}

\def\reals{ { {\rm  I \kern-0.15em R }  } }
\def\complex{ {\,{{\rm C} \kern-0.50em \raise0.20ex {  |}}\, }}

\def\mubf{\hbox{\boldmath$\mu$\unboldmath}}

\def\Qbf{{\bf Q}}
\def\Rbf{{\bf R}}

\def\Xbf{{\bf X}}

\def\Ic{{\cal I}}

\def\Kc{{\cal K}}

\def\Nc{{\cal N}}

\def\be{\vskip .3cm \begin{equation}}
\def\ee{\end{equation} \vskip .4cm \noindent}
\def\defeq{{\stackrel{\Delta}{=}}}
\def\Rxx{\Rbf_{\ssstyle X\kern-.1em X}}

\let\ssstyle=\scriptscriptstyle

\def\Kout{\setbox1=\hbox{\Huge\bf K}\hbox to
1.05\wd1{\hspace{.05\wd1}
}}
\def\Sout{\setbox1=\hbox{\Huge\bf S}\hbox to 1.05\wd1{\hspace{.05\wd1}
}}

\input setup

\usepackage{spconf,amsmath,epsfig,epsf,psfrag,amssymb,amsfonts,latexsym, amsmath,color}
\usepackage{verbatim}
\usepackage[mathscr]{eucal}

\newcommand{\Kmsc}{\mbox{$\mathscr{K}$}}

\newcommand{\beq}{\begin{equation}}
\newcommand{\eeq}{\end{equation}}

\def\defeq{\stackrel{\Delta}{=}}
\def\Ebb{{\mathbb E}}

\definecolor{bgrd}{rgb}{1,1,1}
\definecolor{grey}{rgb}{0.9,0.9,0.6}
\definecolor{gray}{rgb}{0.5,0.5,0.5}

\title{Information, Energy and Density for  {\em Ad Hoc} Sensor Networks over Correlated Random
Fields: Large Deviations Analysis}
\name{Youngchul Sung\sthanks{{\scriptsize Y. Sung and H. Yu are
with the Dept. of Electrical Engineering, KAIST, Daejeon 305-701,
South Korea.  Email:ysung@ee.kaist.ac.kr and
hjyu@stein.kaist.ac.kr. H. V. Poor is with the Dept. of Electrical
Engineering,
  Princeton University, Princeton, NJ 08544. Email: poor@princeton.edu. The work of Y. Sung was  supported in
part by Brain Korea 21 Project, the School of Information
Technology, KAIST. The work of H. V. Poor was supported in part by
the U. S. National Science Foundation under Grants ANI-03-38807
and CNS-06-25637.}}, H. Vincent Poor and Heejung Yu }
\address{}

\begin{document}

\maketitle

\begin{abstract}
Using large deviations results that characterize the amount of
information
 per node on a two-dimensional (2-D) lattice, asymptotic behavior of a sensor
network deployed over a correlated random field for statistical
inference is investigated. Under a 2-D hidden Gauss-Markov
random field model with symmetric first order conditional
autoregression, the behavior of the total information [nats] and
energy efficiency [nats/J] defined as the ratio of total gathered
information to the required energy is obtained as the coverage
area, node density and energy vary.
\end{abstract}
% no key words

\vspace{-0.5em}
\section{Introduction}
\vspace{-0.3em}

 In this paper, we investigate the
fundamental behavior of a  flat multi-hop {\em ad hoc} sensor
network deployed over a correlated two-dimensional (2-D) random
field for statistical inference. In particular, we examine the
amount of information obtainable from a sensor network distributed
over a 2-D Gauss-Markov random field (GMRF) and related trade-offs
in various asymptotic settings. We consider the Kullback-Leibler
information (KLI) and mutual information (MI)
\cite{Liese&Vajda:06IT} as our information measures.  Our approach
to calculating the total obtainable information is based on the
large deviations principle. That is, for large networks the total
information  is approximately given by the product of the number
of sensors and the asymptotic per-sensor information.  However, a
closed-form expression for the asymptotic per-sensor information
(or asymptotic information rate in 2-D) is not available for
general 2-D signals. To address this problem, we adopt the {\em
conditional autoregression (CAR) model} and corresponding
correlation model for the signal, and derive a closed-form
expression for the asymptotic information rate in 2-D.
 We do so by exploiting the spectral structure of the CAR signal and the relationship between the eigenvalues of the block circulant
 approximation to a block Toeplitz matrix describing the 2-D correlation
 structure. Based on the derived expressions for asymptotic information rate
and their properties, we investigate the  behavior of sensor
networks deployed over correlated random fields for statistical inference.

\vspace{-0.5em}
%%%%%%%%%%%%%%%%%%%%%%%%%%%%%%%%%
\subsection{Related Work}
%%%%%%%%%%%%%%%%%%%%%%%%%%%%%%%%%
\vspace{-0.3em}

 Large deviations analysis of Gauss-Markov
processes in Gaussian noise has been considered previously.  (See
\cite{Sung&Tong&Poor:06IT} and references therein.) However, most
work in this area considers only one-dimensional (1-D) signals or
time series. A closed-form expression for the asymptotic KLI rate
was obtained and its properties were investigated for 1-D hidden
Gauss-Markov random processes \cite{Sung&Tong&Poor:06IT}. Large
deviations analyses were used to examine the  issues of optimal
sensor density  and optimal sampling in a 1-D signal model in
\cite{Sung&Zhang&Tong&Poor:08SP} and
\cite{Chamberland&Veeravalli:06IT}. For a 2-D setting, an error
exponent was obtained for the detection of 2-D GMRFs in
\cite{Anandkumar&Tong&Swami:07ICASSP}, where the sensors are
located randomly and the Markov graph is based on the nearest
neighbor  dependency enabling a loop-free graph. In this work,
however, measurement noise was not considered.   Our work here
focuses on the analysis of the fundamental behavior of 2-D sensor
networks deployed for statistical inference via new large
deviations results for 2-D {\it hidden} GMRFs, which enable us to
investigate the impact of field correlation and measurement
signal-to-noise ratio (SNR) on the information.

\vspace{-0.7em}
%%%%%%%%%%%%%%%%%%%%%%%%%%%%%%%%%%%%%%%%%%%%%%%%%%%%%%%%%%%%%%%%%%%%%%%
\section{Background and Signal Model}
\label{sec:systemmodel}

\begin{figure}[htbp]
\centerline{
    \begin{psfrags}
    \psfrag{ij}[l]{{\scriptsize $(i,j)$}}
    \psfrag{xij}[c]{{\scriptsize $X_{ij}$}}
    \psfrag{wij}[l]{{\scriptsize $W_{ij}$}}
    \psfrag{yij}[c]{{\scriptsize $Y_{ij}$}}
    \psfrag{Nij}[c]{{\scriptsize Sensor $ij$}}
    \psfrag{r}[c]{{\scriptsize $d_n$}}
    \scalefig{0.33}\epsfbox{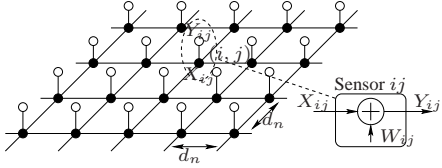}
    \end{psfrags}
} \caption{Sensors on a 2-D Lattice ${\mathcal I}_n$: Hidden Markov
Structure} \label{fig:2dHGMRF}
\end{figure}

To simplify the problem and gain insights into behavior in 2-D, we
assume that sensors are located on a 2-D lattice
$\Ic_n=[0:1:n-1]^2$, as shown in Fig. \ref{fig:2dHGMRF}. We assume
that the signal samples of sensors form a (discrete-index) 2-D
GMRF and that each sensor has Gaussian measurement noise. The
(noisy) measurement $Y_{ij}$ of Sensor $ij$ on the 2-D lattice
${\mathcal I}_n$ is given by
\begin{equation} \label{eq:hypothesis2d}
 Y_{ij} = X_{ij}+ W_{ij}, ~~ij \in {\cal I}_n,
\end{equation}
where  $\{W_{ij}\}$ represents independent and identically
distributed (i.i.d.) $\Nc(0,\sigma^2)$ noise with a known variance
$\sigma^2$, and $\{X_{ij}\}$ is a  GMRF on the 2-D lattice
independent of the measurement noise $\{W_{ij}\}$. Thus, the
observation samples form a 2-D hidden GMRF. In the following, we
briefly introduce the results on GMRFs relevant to further
development.

\vspace{0.1em}
\begin{definition}[GMRF \cite{Rue&Held:book}]\label{def:GMRF}
A random vector $\Xbf=(X_1,X_2,$ $\cdots,X_n)$ $\in {\mathbb R}^n$
is a Gauss-Markov random field  with respect to (w.r.t.) a
labelled graph ${\mathcal G}=({\mathcal \nu},{\mathcal E})$ with
mean $\mubf$ and precision matrix $\Qbf
>0$, if its probability density function is given by
{\footnotesize
\begin{equation}
p(\Xbf) = (2\pi)^{-n/2}|\Qbf|^{1/2}\exp\left( - \frac{1}{2}
(\Xbf-\mubf)^T \Qbf (\Xbf-\mubf) \right),
\end{equation}}
and $Q_{lm} \ne 0 \Longleftrightarrow \{l,m\} \in {\mathcal
E}~\mbox{for all}~ l \ne m$.  Here, ${\mathcal \nu}$ is  the set
of all nodes $\{1,2,\cdots, n\}$ and ${\mathcal E}$ is the set of
edges connecting pairs of nodes, which represent the conditional
dependence structure.
\end{definition}

 The 2-D
indexing scheme $ij$ in (\ref{eq:hypothesis2d}) can be
appropriately converted to an 1-D scheme to apply Definition
\ref{def:GMRF}. From here on, we use the 2-D indexing scheme for
convenience.

\begin{definition}[Stationarity]
A GMRF $\{X_{ij}\}$  on a 2-D doubly infinite lattice ${\mathcal
I}_\infty$ is said to be {stationary} if the mean vector is
constant and ${\rm Cov}( X_{ij}, X_{i^\prime j^\prime})$ $\defeq
\Ebb \{(X_{ij}-\Ebb\{X_{ij}\})$ $(X_{i^\prime j^\prime}-\Ebb
\{X_{i^\prime j^\prime}  \}) \}= c(i-i^\prime, j-j^\prime)$ ~for
some function $c(\cdot, \cdot)$.
\end{definition}

 For a 2-D stationary GMRF $\{X_{ij}\}$, the
covariance $\{\gamma_{ij}\}$ is defined as $\gamma_{ij} = \Ebb \{
X_{i^\prime j^\prime} X_{i^\prime+i, j^\prime +j}\} =\Ebb \{
X_{00} X_{ij}\},$ which does not depend on $i^\prime$ or
$j^\prime$ due to the stationarity. The spectral density function
of a zero-mean and stationary GMRF  on ${\mathcal I}_\infty$ with
covariance $\gamma_{ij}$ is defined as
\begin{equation}  \label{eq:2DDTFT}
f(\omega_1,\omega_2) =  \frac{1}{4\pi^2}\sum_{ij \in {\mathcal
I}_\infty} \gamma_{ij} \exp(-\iota(i\omega_1 + j\omega_2) ),
\end{equation}
where $\iota = \sqrt{-1}$ and $(\omega_1,\omega_2) \in
(-\pi,\pi]^2$. Note that this is a 2-D extension of the
conventional 1-D discrete-time Fourier transform (DTFT).

\vspace{0.2em}
\begin{definition}[The Conditional Autoregression ]
A GMRF $\{X_{ij}\}$ is said to be a conditional autoregression
(CAR) if it is specified using a set of full conditional normal
distributions with mean and precision: {\footnotesize
\begin{eqnarray}
\Ebb \{ X_{ij}|\Xbf_{-ij}\} &=&  -\frac{1}{\theta_{00}}
\sum_{i^\prime j^\prime \in {\mathcal I}_\infty \ne 00}
\theta_{i^\prime j^\prime} X_{i+i^\prime,j+j^\prime}, \label{eq:condMean2DInf}\\
\mbox{Prec}\{X_{ij}|\Xbf_{-ij}\} &=& \theta_{00} > 0,
\label{eq:condPrec2DInf}
\end{eqnarray}}
where $\Xbf_{-ij}$ denotes the set of all variables except
$X_{ij}$.
\end{definition}
It is shown that the GMRF defined by the CAR model
(\ref{eq:condMean2DInf}) - (\ref{eq:condPrec2DInf}) is a zero-mean
stationary Gaussian process on ${\mathcal I}_\infty$ with the power
spectral density \cite{Rue&Held:book}
\begin{equation}
f(\omega_1,\omega_2) =  \frac{1}{4\pi^2} \frac{1}{\sum_{ij \in
{\mathcal I}_\infty} \theta_{ij} \exp(-\iota (i\omega_1 +
j\omega_2))}
\end{equation} {\footnotesize
\begin{eqnarray}
\mbox{if}&&|\{\theta_{ij} \ne 0\}| < \infty, ~~~~ \theta_{ij} = \theta_{-i,-j}, ~~~~ \theta_{00} >0, \label{eq:CARcond1}\\
&&\{\theta_{ij}\} ~\mbox{is so that}~ f(\omega_1,\omega_2)>0, ~~~
\forall (\omega_1,\omega_2) \in (-\pi,\pi]^2. \label{eq:CARcond4}
\end{eqnarray}}
Henceforth, we assume that the 2-D stochastic signal $\{X_{ij}\}$ in
(\ref{eq:hypothesis2d}) is given by a stationary GMRF defined by
the CAR model (\ref{eq:condMean2DInf}) - (\ref{eq:condPrec2DInf})
and (\ref{eq:CARcond1}) -  (\ref{eq:CARcond4}).

\vspace{-0.7em}
%%%%%%%%%%%%%%%%%%%%%%%%%%%%%%%%%%%%%%%%%%%%%%%%%%%%%%%%%%%%%%%%%%%
\section{Asymptotic Information Rates and Their Properties}
%%%%%%%%%%%%%%%%%%%%%%%%%%%%%%%%%%%%%%%%%%%%%%%%%%%%%%%%%%%%%%%%%%%
\vspace{-0.2em}

In this section, we derive a closed-form expression for the
asymptotic KLI rate and MI rate in the model (\ref{eq:hypothesis2d}), defined as
\[
\Kmsc = \lim_{n \rightarrow \infty} \frac{1}{|\Ic_n|} \log
\frac{p_{0}}{p_{1}}(\{Y_{ij}, ij \in \Ic_n\}) ~\mbox{a.s.
under}~p_{0}, ~~\mbox{and}
\]
\[
I = \lim_{n \rightarrow \infty} \frac{1}{|\Ic_n|} I(\{X_{ij}, ij
\in \Ic_n\};\{Y_{ij}, ij\in \Ic_n\}),~~~~~~~~~
\]
respectively. For the MI, the signal model (\ref{eq:hypothesis2d})
is directly applicable, whereas for the KLI the probability density functions of the
null (noise-only) and
alternative (signal-plus-noise) distributions are given by
\begin{eqnarray}
p_0 (Y_{ij}) &:&  Y_{ij} = W_{ij} , ~~ij \in {\cal I}_n, \nonumber\\
p_1(Y_{ij})  &:& Y_{ij} = X_{ij}+ W_{ij}, ~~ij \in {\cal I}_n.
\label{eq:KLIp0p1}
\end{eqnarray}
The following closed-form expressions for the asymptotic information
rates in the spectral domain have been obtained in \cite{Sung&Poor&Yu:08ICASSP}
by exploiting the spectral structure
of the CAR signal and the relationship between the eigenvalues of
block circulant and block Toeplitz matrices representing 2-D
correlation structure.

\vspace{0.3em}
\begin{theorem}\label{theo:AKLIR_CAR}  For the model (\ref{eq:KLIp0p1}) with the signal given by
(\ref{eq:condMean2DInf}) - (\ref{eq:condPrec2DInf}),  assuming
that conditions
 (\ref{eq:CARcond1}) -  (\ref{eq:CARcond4}) hold, the asymptotic KLI rate is
given by {\scriptsize
\begin{eqnarray}
\Kmsc &=& \frac{1}{4\pi^2} \int_{-\pi}^{\pi} \int_{-\pi}^{\pi}
\biggl( \frac{1}{2}\log
\frac{\sigma^2+4\pi^2f(\omega_1,\omega_2)}{\sigma^2}\label{eq:errorexponentspectral}\\
&& ~~~~~~~~~~~~~~~~~~~~~~~~~~~~~~~~ +\frac{1}{2}
\frac{\sigma^2}{\sigma^2+4\pi^2f(\omega_1,\omega_2)} -\frac{1}{2}
\biggr)d\omega_1d\omega_2, \nonumber \\
&=&\frac{1}{4\pi^2}\int_{-\pi}^{\pi} \int_{-\pi}^{\pi} D(
\Nc(0,S_{0}^y(\omega_1,\omega_2))||\Nc(0,S_{1}^y(\omega_1,\omega_2))
~d\omega_1 d\omega_2,\nonumber
\end{eqnarray}
} where $D(\cdot||\cdot)$ denotes the Kullback-Leibler divergence.
\end{theorem}
{\em Proof:} In \cite{Sung&Poor&Yu:08ITsub}.

\vspace{0.3em} As a by-product of the proof of the above theorem,
we have the asymptotic MI rate given by
\begin{equation}
I = \frac{1}{4\pi^2} \int_{-\pi}^{\pi} \int_{-\pi}^{\pi}
\frac{1}{2}\log
\frac{\sigma^2+4\pi^2f(\omega_1,\omega_2)}{\sigma^2}
  d\omega_1d\omega_2.\label{eq:persensormutualinformation}\\
\end{equation}
Theorem \ref{theo:AKLIR_CAR} is a 2-D extension of the asymptotic
KLI rate  of 1-D hidden Gauss-Markov model  obtained in
\cite{Sung&Tong&Poor:06IT}, and the asymptotic KLI rate
(\ref{eq:errorexponentspectral}) can be explained using a
frequency binning argument. Specifically, for each 2-D frequency bin $d\omega_1
d\omega_2$, the spectra are flat, i.e., the signals are
independent and Stein's lemma can be applied for the bin. The
overall KLI  is the sum of contributions from each segment.

\vspace{-0.5em}
%%%%%%%%%%%%%%%%%%%%%%%%%%%%%%%%%%%%%%%%%%%%%%%%%%%%%%%%%%%%%%%%%%
\subsection{Symmetric First Order Conditional Autoregression}

To investigate the properties of the asymptotic KLI and MI rates
as functions of field correlation and SNR, we further consider
the symmetric first order conditional autoregression (SFCAR),
defined by the conditions {\small
\begin{eqnarray*}
\Ebb \{ X_{ij}|\Xbf_{-ij}\} &=&  \frac{\lambda}{\kappa} (X_{i+1,j}+X_{i-1,j}+X_{i,j+1}+X_{i,j-1}),\\
\mbox{Prec}\{X_{ij}|\Xbf_{-ij}\} &=& \kappa > 0,
\end{eqnarray*}}
where $0 \le \lambda \le \frac{\kappa}{4}$. (This is a sufficient
condition to satisfy (\ref{eq:CARcond1}) - (\ref{eq:CARcond4}).)
Here, $\theta_{00}=\kappa$ and $\theta_{1,0} = \theta_{-1,0} =
\theta_{0,1} = \theta_{0,-1} = -\lambda$. In the SFCAR model, the
correlation is symmetric for each set of four neighboring sensor
nodes. The SFCAR model is a simple but meaningful extension of the
1-D autoregression (AR) model which has the conditional causal
dependency only on the previous sample. Here in the 2-D case we
have conditional dependence on four neighboring nodes in the four
(planar) directions, capturing 2-D correlation structure. The
spectrum of the SFCAR signal is given by
\begin{equation}
f(\omega_1,\omega_2) = \frac{1}{4\pi^2 \kappa (1 - 2 \zeta
\cos\omega_1 - 2 \zeta \cos\omega_2)},
\end{equation}
where the {\em edge dependence factor} $\zeta$ is defined as
\vspace{-0.5em}
\begin{equation}
\zeta
\defeq \frac{\lambda}{\kappa}, ~~~~ 0 \le \zeta \le 1/4.
\end{equation}
Here, $\zeta =0$ corresponds to the i.i.d. case whereas $\zeta
=1/4$ corresponds to the perfectly correlated case. Therefore, the
correlation strength can be captured in this single quantity
$\zeta$ for SFCAR signals. The power of the SFCAR is obtained
using the inverse Fourier transform via the relationship
(\ref{eq:2DDTFT}), and is given by
%\begin{equation}
$P_s = \gamma_{00} = \frac{2K(4\zeta)}{\pi \kappa}, ~\left(0 \le
\zeta \le \frac{1}{4} \right)$,
%\end{equation}
where $K(\cdot)$ is
the complete elliptic integral of the first kind
\cite{Besag:81JRSS}. The SNR is given by
%\begin{equation} \label{eq:SNR}
$ \mbox{SNR} = \frac{P_s}{\sigma^2} = \frac{2K(4\zeta)}{\pi \kappa
\sigma^2}$.
%\end{equation}
Using  (\ref{eq:errorexponentspectral}) and the SNR, we obtain
the asymptotic KLI and MI rates for the SFCAR signal, given in the
following corollary to Theorem 1, also from \cite{Sung&Poor&Yu:08ICASSP}.

\vspace{0.3em}
\begin{corollary}\label{corol:eeSFA}
The asymptotic KLI and MI rates for the SFCAR 2D signal model are
given by {\tiny
\begin{eqnarray}
\Kmsc_s &=& \frac{1}{4\pi^2} \int_{-\pi}^{\pi} \int_{-\pi}^{\pi}
\biggl( \frac{1}{2}\log \left(1+\frac{ \mbox{SNR}}{
(2/\pi)K(4\zeta) (1 - 2 \zeta \cos\omega_1 - 2 \zeta
\cos\omega_2)}\right) \nonumber\\
&& ~~~~~~+\frac{1}{2} \frac{1}{1+\frac{ \mbox{SNR}}{
(2/\pi)K(4\zeta) (1 - 2 \zeta \cos\omega_1 - 2 \zeta
\cos\omega_2)}} -\frac{1}{2} \biggl)d\omega_1d\omega_2.
\label{eq:aKLIR_SFA}
\end{eqnarray}
} and {\tiny
\begin{equation}
I_s = \frac{1}{4\pi^2} \int_{-\pi}^{\pi} \int_{-\pi}^{\pi}
\frac{1}{2}\log \left(1+\frac{ \mbox{SNR}}{ (2/\pi)K(4\zeta) (1 -
2 \zeta \cos\omega_1 - 2 \zeta \cos\omega_2)}\right)
d\omega_1d\omega_2, \label{eq:aMLIR_SFA}
\end{equation}
} respectively.
\end{corollary}
\vspace{0.3em}

Note that the SNR and correlation are separated in
(\ref{eq:aKLIR_SFA})-(\ref{eq:aMLIR_SFA}), which enables us to
investigate the effects of each term separately.

\vspace{-0.5em}
%%%%%%%%%%%%%%%%%%%%%%%%%%%%%%%%%%%%%%%%%%%%%%%%%%%%%%%%%%%%%%%%%%%%
\subsection{Properties of the asymptotic KLI and MI rates ($\Kmsc_s$ and $I_s$)
}\label{subsec:AIRSFAproperties}

First, it is readily seen from Corollary \ref{corol:eeSFA} that
$\Kmsc_s$ and $I_s$ are continuously differentiable $C^1$
functions of the edge dependence factor $\zeta$ ($0 \le \zeta \le
1/4$) for a given SNR since $f:x \rightarrow K(x)$ is a
continuously differentiable $C^\infty$ function for $0 \le x < 1$
\cite{Erdelyi:53book}. The values of $\Kc_s$ at the extreme
correlations are given by noting that $K(0) = \frac{\pi}{2} \ \
{\rm and} \ \ K(1)= \infty$. Therefore, in the i.i.d. case ($\zeta
=0$), the corollary reduces to Stein's lemma as expected, and
$\Kmsc_s$ is given by {\scriptsize
\[
\Kmsc_s|_{\zeta=0}= \frac{1}{2} \log (1+ \mbox{SNR})
+\frac{1}{2(1+ \mbox{SNR})} -\frac{1}{2}
=D(\Nc(0,1)||\Nc(0,1+\mbox{SNR})).
\]
} In the i.i.d. case,  the asymptotic MI rate is given by the well
known formula, $I_s|_{\zeta=0} = \frac{1}{2} \log (1 +
\mbox{SNR})$. For the perfectly correlated case ($\zeta=1/4$), on
the other hand, $\Kmsc_s =0$ and $I_s =0$. (In this case as well
as in the i.i.d. case, the two-dimensionality is irrelevant.) The
limiting behavior of the asymptotic information rates is given by
Taylor's theorem. Due to the continuous differentiability, we have
\begin{eqnarray}
\Kmsc_s (\zeta) &=& c_1 \cdot (1/4-\zeta) +
o ( |1/4-\zeta | ), \label{eq:KmscQuaterZeta} \\
I_s (\zeta) &=&  c_1^\prime \cdot (1/4 - \zeta ) + o ( |1/4 -
\zeta | ),\label{eq:KmscQuaterZeta2}
\end{eqnarray}
for some constants $c_1$ and $c_1^\prime$, as $\zeta \rightarrow
1/4$. Similarly, we also have the linear limiting behavior for
$\Kmsc_s$ and $I_s$ in a neighborhood of $\zeta =0$ with non-zero
limit values, as $\zeta \rightarrow 0$. That is,
\begin{eqnarray}
\Kmsc_s(\zeta) &=& \Kmsc_s(0) + c_2 \zeta + o(\zeta),\label{eq:Kmsc0Zeta}\\
I_s(\zeta) &=& I_s(0) + c_2^\prime \zeta +
o(\zeta),\label{eq:Kmsc0Zeta2}
\end{eqnarray}
for some $c_2$ and $c_2^\prime$, as $\zeta \rightarrow 0$. For
intermediate values of correlation, it is seen that at high SNR
$\Kmsc_s$ is monotonically decreasing as $\zeta$ increases. At low
SNR, on the other hand, correlation is beneficial to the
performance.

 With regard to $\Kmsc_s$  and $I_s$ as functions of SNR,
 the behavior of $\Kmsc_s$ is given by the following theorem.

\vspace{0.5em}
\begin{theorem} \label{theo:KLIsvsSNR}
The asymptotic KLI rate  $\Kmsc_s$ for the hidden SFCAR model is
continuous and monotonically increasing as SNR increases for a
given edge dependence factor $0 \le \zeta < 1/4$. Moreover,
$\Kmsc_s$ increases linearly with respect to $\frac{1}{2}\log
\mbox{SNR}$ as $\mbox{SNR} \rightarrow \infty$. As SNR decreases
to zero, on the other hand, $\Kmsc_s$ converges to zero with the
convergence rate $\Kmsc_s(\mbox{SNR}) = c_3\cdot \mbox{SNR}^2  +
o(\mbox{SNR}^2)$ for some constant $c_3$ as $\mbox{SNR}
\rightarrow 0$.
 The asymptotic MI rate  $I_s$
has similar properties as a function of SNR, i.e., it is a
continuous and monotonically-increasing function of SNR. At high
SNR, it increases with rate $\frac{1}{2}\log \mbox{SNR}$, whereas
it decreases to zero with rate of convergence $I_s(\mbox{SNR}) =
c_3^\prime \cdot\mbox{SNR} + o(\mbox{SNR})$ for some constant
$c_3^\prime$ as $\mbox{SNR} \rightarrow 0$.
\end{theorem}

 {\em Proof:} In \cite{Sung&Poor&Yu:08ITsub}.

Note that the limiting behavior as $\mbox{SNR} \rightarrow 0$ is
different for $\Kmsc_s$ and $I_s$; $\Kmsc_s$ decays to zero
quadratically while $I_s$ diminishes linearly.

%%%%%%%%%%%%%%%%%%%%%%%%%%%%%%%%%%%%%%%%%%%%%%%%%%%%%%%%%%%%%%%%%%%%%%%%
\section{Scaling Laws in Ad Hoc Sensor Networks over Correlated Random Field}
\label{sec:adHocFundBehave} %%%%%%%%%%%%%%%%%%%%%%%%%%%%%%%%%%%%%%%%%%%%

Based on the  results in the previous sections, we are now ready
to answer some fundamental questions in the design of sensor
networks for statistical inference about the underlying stochastic
field.

\vspace{-0.5em}
\subsection{Physical correlation model}
\label{subsec:physicalmodel}

The actual physical correlation for the SFCAR  model is given by
solving the corresponding continuous-index 2-D stochastic
differential equation (the stochastic Laplace
equation)\footnote{Note that the solution of (\ref{eq:laplaceSDE})
is circularly symmetric, i.e., it depends only on
$r=\sqrt{x^2+y^2}$, and samples of the solution
 $X(x,y)$ of (\ref{eq:laplaceSDE}) on lattice $\Ic_n$ do not
necessarily form a discrete-index SFCAR GMRF. However,
(\ref{eq:laplaceSDE}) is still the continuous-index counterpart of
the SFCAR model, and we use its correlation function for the SFCAR
model.} \cite{Whittle:54Biometrika} {\footnotesize
\begin{equation} \label{eq:laplaceSDE}
\left[ \left( \frac{\partial}{\partial x}\right)^2 +\left(
\frac{\partial}{\partial y} \right)^2 - \alpha^2 \right]X(x,y) =
u(x,y),
\end{equation}} where $u(x,y)$ is the 2-D white zero-mean
Gaussian perturbation and $\alpha > 0$ is the diffusion rate. By
solving the SDE, the {\em edge correlation factor} $\rho$ is
given, as a function of the sensor spacing $d_n$, by
\cite{Whittle:54Biometrika}
\begin{equation} \label{eq:2DcorrelationFunc}
\rho \defeq
\frac{\gamma_{01}}{\gamma_{00}}=\frac{\gamma_{10}}{\gamma_{00}}=
f(d_n) = \alpha d_n K_1(\alpha d_n),
\end{equation}
where $K_1(\cdot)$ is the modified Bessel function of the second
kind whose asymptotic behavior is given by
\begin{equation} \label{eq:modifiedBessel}
\left\{
\begin{array}{cccl}
K_1(x) &\rightarrow& \sqrt{\frac{\pi}{2x}}e^{-x} & \mbox{as}~ x
\rightarrow \infty,\\
K_1(x) &\rightarrow& 1/x & \mbox{as}~ x
\rightarrow 0.\\
\end{array}
\right.
\end{equation}
 The correlation function (\ref{eq:2DcorrelationFunc}) can be regarded
as the representative  correlation in 2-D, similar to the
exponential correlation function $e^{-Ad_n}$ in 1-D. Both
functions decrease monotonically w.r.t. $d_n$. However, the 2-D
correlation function is flat at $d_n=0$
\cite{Whittle:54Biometrika}. Further, we have a continuous and
differentiable mapping $g:\rho \rightarrow \zeta$ from the edge
correlation factor $\rho$ to the edge dependence factor $\zeta$,
given by \cite{Sung&Poor&Yu:08ITsub}
\begin{equation} \label{eq:ZetaVsRho}
\rho =  \frac{(2/\pi)K(4\zeta)-1}{4 (2/\pi)\zeta K(4\zeta)} =:
g^{-1}(\zeta),
\end{equation}
which maps zero and one to zero and 1/4, respectively. Thus, we
have $\zeta = g(f (d_n)),$ and for given physical parameters (with
a slight abuse of notation),
\[
\Kmsc_s(\mbox{SNR},\zeta) = \Kmsc_s(\mbox{SNR},g(f(d_n))) =
\Kmsc_s(\mbox{SNR},d_n).
\]
(And, similarly for $I_s$.) We will use the arguments SNR and $\zeta$
for $\Kmsc_s$ and $I_s$ properly if necessary.

\vspace{-0.4em}
%%%%%%%%%%%%%%%%%%%%%%%%%%%%%%%%%%%%%%%%%%%%%%%%%%%%%%%%%%%%%%%%%%%
\subsection{Asymptotic behavior}
\label{subsec:adhocfundamentalbehave}

In the following, we summarize the assumptions for the planar {\em
ad hoc} sensor network that we consider.
\begin{itemize}
\item[(A.1)]    $n^2$ sensors are located on the grid $\Ic_n =
[0:1:n-1]^2$ with spacing $d_n$, as shown in Fig.
\ref{fig:2dHGMRF}, and a fusion center is located
at the center $(\lfloor n/2\rfloor,\lfloor n/2\rfloor)$.

\item[(A.2)]
The observations $\{Y_{ij}\}$ at sensor nodes form a 2-D hidden (discrete-index) SFCAR Gauss-Markov random field on the lattice for each $d_n >0$,
and the edge dependence factor is given by (\ref{eq:2DcorrelationFunc}) and (\ref{eq:ZetaVsRho}).%

\item[(A.3)] The fusion center gathers the measurement from all
nodes using the minimum hop routing. Note that the links in Fig.
\ref{fig:2dHGMRF} are not only the Markov dependence edges but
also the routing links.  The minimum hop routing requires a hop
count of $|i-\lfloor n/2\rfloor|+|j-\lfloor n/2\rfloor|$ to
deliver $Y_{ij}$ to the fusion center.

\item[(A.4)] The communication  energy per link $E_{c}(d_n) =  E_0
d_n^{\nu}$, where  $\nu \ge 2$ is the propagation loss factor in
wireless channel.

\item[(A.5)] Sensing requires energy, and the sensing energy per
node is denoted by $E_{s}$.  Moreover, we assume that the {\em
measurement} SNR  increases linearly w.r.t. $E_{s}$, i.e.,
$\mbox{SNR} = \beta E_s$  for some constant $\beta$.
\end{itemize}

\noindent  Henceforth, we consider various asymptotic scenarios
and investigate the fundamental behavior of the {\em ad hoc}
sensor network deployed over a correlated random field for
statistical inference under assumptions
\textit{(A.1)}-\textit{(A.5)}. (Proofs are omitted due to limited
space.)

 The sensor density $\mu_n$ on
$\Ic_n$ is given by $\mu_n = \frac{n^2}{((n-1)d_n)^2}$. Assuming
that the network is sufficiently large, the total information
about the underlying field obtainable from the network is given by
\begin{equation} \label{eq:adhocTotalInfo}
\mbox{KLI}_T = n^2 \Kmsc_s ~\mbox{and}~\mbox{MI}_T = n^2 I_s,
\end{equation}
and the total consumed energy in the network is given by
\begin{eqnarray}
E &=& n^2 E_s + E_c(d_n) \sum_{i=0}^{n-1}\sum_{j=0}^{n-1}
(|i-\lfloor n/2 \rfloor|+|j-\lfloor n/2 \rfloor|),\nonumber\\
&=& n^2 E_s + \Theta(n^3) E_{c}(d_n).
\label{eq:totalconsumedenergy}
\end{eqnarray}
Note that the knowledge of per-node information $\Kmsc_s$ and
$I_s$ and their properties w.r.t. SNR and sensor spacing $d_n$ in
(\ref{eq:adhocTotalInfo}) is critical for further development, and
it is provided in the previous sections.

We begin with the increasing area case.

\vspace{0.2em}
\begin{theorem}[Infinite area and fixed density] For an {\em ad hoc} sensor network
with  a fixed and finite node density, the total amount of
information increases linearly as the area increases, but under
both information measures the amount of harvested information per
unit energy decays to zero with rate
\begin{equation} \label{eq:efficiencyIAM}
\eta = \Theta\left(\mbox{area}^{-1/2} \right),
\end{equation}
for any non-trivial diffusion rate $\alpha$, i.e., $0 < \alpha <
\infty$ as we increase the area.
\end{theorem}
\vspace{0.2em}

 Next, we consider the case in which the node density
diminishes, i.e., $d_n \rightarrow \infty$.  This case
is of particular interest at high SNR since at high SNR less correlated
samples yield larger per-node information. However, the per-sensor
information is upper bounded as $d_n \rightarrow \infty$, and the
asymptotic behavior is given by the following theorem.

\vspace{0.2em}
\begin{theorem} \label{theo:informationvsdninfty}
As $d_n \rightarrow \infty$, the per-node information $\Kmsc_s$
and $I_s$ converge to
$\Kmsc_s(0)=D(\Nc(0,1)||\Nc(0,1+\mbox{SNR}))$ and
$I_s(0)=\frac{1}{2}\log \mbox{SNR}$, respectively, and the
convergence rate is given by {\footnotesize
\begin{eqnarray}
\Kmsc_s(d_n) &=&   \Kmsc_s(0) - c_4\sqrt{d_n}e^{- \alpha d_n} +
o\left(\sqrt{d_n}e^{-\alpha d_n}\right), \label{eq:theoremdninfty}\\
I_s(d_n) &=& I_s(0)  - c_4^\prime\sqrt{d_n}e^{- \alpha d_n} +
o\left(\sqrt{d_n}e^{-\alpha d_n}\right),
\end{eqnarray}}
for constants $c_4,~c_4^\prime >0$ depending on the SNR.
\end{theorem}
\vspace{0.2em}

\begin{figure}[htbp]
\centerline{
    \begin{psfrags}
    \psfrag{ER}[r]{{ $E(R)$}}  %
    \psfrag{R}[c]{{ $R$}} %
    \psfrag{R0}[c]{{ $R_0$}} %
    \psfrag{Rc}[c]{{ $R_c$}} %
    \psfrag{I1}[c]{ $I$} %
    \psfrag{A}[c]{ $A$}
    \psfrag{I}[c]{{Mutual information}} %
     \scalefig{0.45}\epsfbox{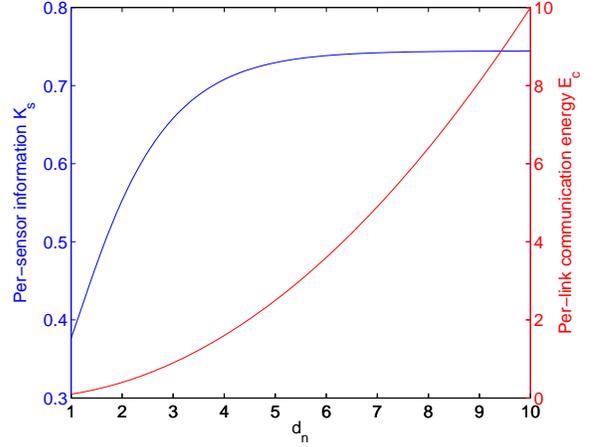}
    \end{psfrags}
} \caption{Per-node information and per-link communication energy
w.r.t. sensor spacing $d_n$ (SNR = 10 dB, $\alpha=1$, $c_4=1$)}
\label{fig:dngoestoinfty}
\end{figure}

Theorem \ref{theo:informationvsdninfty} can be proved using
(\ref{eq:Kmsc0Zeta}, \ref{eq:Kmsc0Zeta2}) and
(\ref{eq:2DcorrelationFunc}, \ref{eq:modifiedBessel}), and
explains how much gain is obtained from less correlated
observations by increasing the sensor spacing in 2-D. Fig.
\ref{fig:dngoestoinfty} shows $\Kmsc_s$ and $E_c$ as functions of
$d_n$ for $\alpha =1$, $c_4=1$ and 10 dB SNR. The gain in
information is given by $\sqrt{d_n}e^{-\alpha d_n}$ for large
$d_n$,  whereas the required per-link communication energy
increases without bound, i.e., $E_c (d_n) = E_0 d_n^\nu$ ~($\nu
\ge 2$). Since the exponential term is dominant in the gain as
$d_n$ increases, the information gain obtained by increasing $d_n$
decreases almost exponentially, and there is no significant gain
by increasing the sensor spacing further after some value. Hence,
it is not effective in terms of energy efficiency to deploy a very
sparse network aiming at less correlated samples at high SNR.

The per-link communication energy can be made arbitrarily small by
decreasing the sensor spacing. To investigate the effect of
diminishing communication energy $E_c$ as $d_n \rightarrow 0$, we
now consider the asymptotic case in which the node density goes to
infinity for a fixed coverage area.  In this case, the per-node
information decays to zero as $d_n \rightarrow 0$ since $\zeta
\rightarrow 1/4$ as $d_n \rightarrow 0$, and $\Kmsc_s(\zeta)$ and
$I_s(\zeta)$ converge to zero as $\zeta \rightarrow 1/4$, as shown
in Section \ref{subsec:AIRSFAproperties}.  The asymptotic behavior
in this case is given by the following theorem.

\vspace{0.2em}
\begin{theorem}[Infinite density model]
\label{theo:infinitedensitymodel} For the infinite density model
with a fixed coverage area, the per-node information decays to
zero with rate
\begin{equation} \label{eq:infintedensityKs}
\Kmsc_s =   {c_5}{\mu_n^{-1}} + o\left({\mu_n^{-1}}\right),
\end{equation}
for some constant $c_5$ as the node density $\mu_n \rightarrow
\infty$. Hence, the amount of total information per unit area
(nats/$m^2$) converges to the constant $c_5$ as $\mu_n \rightarrow
\infty$. Furthermore, in the case of no sensing energy, a non-zero
energy efficiency $\eta$ is achievable if the propagation loss
factor $\nu =3$, and even an infinite energy efficiency is
achievable if $\nu > 3$ as $\mu_n \rightarrow \infty$ for fixed
area.\footnote{Of course, this depends on the assumption of
$E_c(d_n) = E_0 d_n^\nu$ for any $d_n >0$. However, this
assumption may not be valid for small $d_n$.}
\end{theorem}
\vspace{0.2em}

 The finite total information for the infinite
density and fixed area model follows our intuition.  The maximum
information provided by the samples from the continuous-index
random field does not exceed the information between $X(x,y)$ and
$Y(x,y)$ except for the case of spatially white fields. It is common
that the propagation loss factor $\nu
> 3$ for near field propagation (i.e., $d_n \rightarrow 0$). Hence, infinite energy efficiency
is achievable as we increases the node density for a fixed area
considering only communication energy.  Note that the total amount
of information converges to a constant as we increases the node
density. So, the infinite energy efficiency is achieved by
diminishing communication energy as $d_n \rightarrow 0$.
Considering the sensing energy, however, infinite energy
efficiency is not feasible since we have in this case
\begin{equation}
E = n^2 E_s + \Theta( n^{3-\nu}) ~\mbox{and}~ \eta = \frac{c_5 +
o(1)}{n^2 E_s + \Theta( n^{3-\nu}) }, ~~~\nu \ge 2,
\end{equation}
as $n\rightarrow \infty$ for fixed coverage area. In this case the
sensing energy $n^2 E_s$ is the dominant factor for low energy
efficiency, and the energy efficiency decreases to zero with rate
$O\left(\mu_n^{-1}\right)$. Thus,  it is critical for a densely
deployed sensor network to minimize the sensing energy or
processing energy for each sensor.

\vspace{0.5em} In the infinite density model, we have observed
that energy is an important factor in efficiency. Now we
investigate the change of total information w.r.t. energy. We fix
the node density and consider two scenarios to increase the
required energy: One is to fix the coverage area also and increase
the sensing energy, and the other is to fix the sensing energy and
increase the coverage area. We assume that the network size is
sufficiently large so that our asymptotic analysis is valid. The
energy asymptotic behavior for two scenarios is summarized in the
following theorem.

\begin{theorem} \label{theo:energyasymptotic}
 As we increase the total energy $E$ consumed by a sensor
network with a fixed node density and fixed area, the total
information increases with rate
\begin{equation} \label{eq:energyAsymptotic1}
\mbox{Total information} = O\left(  \log E \right)
\end{equation}
as $E\rightarrow \infty$. When the node density and sensing energy
are fixed and the increasing energy is used to enlarge the
coverage area, on the other hand, the total amount of information
increases with rate of
\begin{equation} \label{eq:energyAsymptotic2}
\mbox{Total information} = \Theta \left( E^{2/3} \right),
\end{equation}
for any $\nu > 0$, as $E \rightarrow \infty$.
\end{theorem}

\begin{comment}
{\em Proof:} First note that
\[
E = (2n+1)^2 E_s + 2n(n+1)(2n+1)E_c(d_n).
\]
In the first case, $n$ and $d_n$ are fixed, and Theorem
\ref{theo:KLIsvsSNR} is directly applicable. Since the number of
nodes and communication energy are fixed, the sensing energy
increases linearly with the total energy $E$. By Assumption {\em
(A.5)}, the measurement SNR increases linearly with the sensing
energy. Applying Theorem \ref{theo:KLIsvsSNR} yields
(\ref{eq:energyAsymptotic1}).

In the second case, $d_n$ and $E_s$ are fixed, and so is
$E_c(d_n)$.  Hence, we have $E = \Theta(n^3)$. The total
information in this case is given by $(2n+1)^2
\Kmsc_s(\mbox{SNR},d_n)$. Since $\Kmsc_s$ is fixed, the total
information is $\Theta(n^2)$ as $n\rightarrow \infty$, and we have
(\ref{eq:energyAsymptotic2}). \hfill{$\blacksquare$}
\end{comment}

Theorem \ref{theo:energyasymptotic} suggests a guideline for investing
the excess energy.  It is not efficient to invest energy to
improve the quality of sensed samples from a limited area. This
only provides the increase in total information in logarithmic
scale. Rather the energy should be spent to increase the number of
samples by enlarging  the coverage area even if it yields less
accurate samples.

\vspace{-1em} %
%%%%%%%%%%%%%%%%%%%%%%%%%%%%%%%%%%%%%%%%%%%%%%%%%%%%%%%%%%%%%%%%%%%%%
\section{Conclusions}
\label{sec:conclusion} \vspace{-0.5em}

We have analyzed the asymptotic behavior of {ad hoc} sensor
networks deployed over correlated random field for statistical
inference. Using our large deviations results that characterize
the asymptotic information rate in 2-D for GMRFs under the CAR
model, we have obtained fundamental scaling laws for total
information and energy efficiency as the node density, coverage
area and consumed energy change. The results provide guidelines
for sensor network design for statistical inference about 2-D
correlated random fields such as temperature, humidity, density of
a gas on a certain area.

%%%%%%%%%% References %%%%%%%%%%%%%%%%%%%%%%%%%%%%%%%%%%%%%%%%%%%%%%%%%%

%{\scriptsize
%\bibliographystyle{ieeetr}
%\bibliography{referenceBibs} %{IEEEabrv,referenceBibs}
%}

{\scriptsize
\bibliographystyle{plain}

}

\end{document}